\documentclass[amsmath,amssymb,prl,superscriptaddress,11pt]{revtex4}

\usepackage{latexsym}
\usepackage{bm}
\usepackage{graphicx}
\DeclareGraphicsExtensions{.ps}

\begin{document}

\preprint{}

\title{Selective Readout and Back-action Reduction for Wideband \\
Acoustic Gravitational Wave Detectors}

\author{M. Bonaldi}
\email[Corresponding author: ]{bonaldi@science.unitn.it}
\affiliation{Istituto di Fotonica e Nanotecnonolgie CNR-ITC and
INFN Trento, I-38050 Povo (Trento), Italy}
\author{M. Cerdonio }
\affiliation{INFN Padova Section and Department of Physics
University of Padova, via Marzolo 8, I-35100 Padova,
Italy}
\author{L. Conti} \affiliation{INFN Padova Section and
Department of Physics University of Padova, via Marzolo 8, I-35100
Padova, Italy}
\author{M. Pinard}
\affiliation{Laboratoire Kastler Brossel, 4 place Jussieu, F75252
Paris, France }
\author{G.A. Prodi} \affiliation{Department of
Physics University of Trento and INFN Trento, I-38050 Povo
(Trento), Italy}
\author{J.P. Zendri }
\affiliation{INFN Padova Section, via Marzolo 8, I-35100 Padova,
Italy}

\begin{abstract}
We present the concept of selective readout for broadband resonant
mass gravitational wave detectors. This detection scheme is
capable of specifically selecting the signal from the
contributions of the vibrational modes sensitive to the
gravitational waves, and efficiently rejecting the contribution
from non gravitationally sensitive modes. Moreover this readout,
applied to a dual detector, is capable to give an effective
reduction of the back-action noise within the frequency band of
interest. The overall effect is a significant enhancement in the
predicted sensitivity, evaluated at the standard quantum limit for
a dual torus detector. A molybdenum detector, 1 m in diameter and
equipped with a wide area selective readout, would reach spectral
strain sensitivities $\sim \,2\times 10^{-23}$ Hz$^{-1/2}$ between
2-6 kHz.
\end{abstract}


\maketitle

The concept of sensitive and broadband acoustic mass gravitational
wave (GW) detectors has been recently presented with reference to
a dual sphere detector \cite{cerdonio1}. Such a novel class of
detectors may be of great interest, as they would be sensitive in
a broad frequency interval of few kHz in the kHz range, were GW
signals from fully relativistic stellar sources are expected
\cite{sorgenti}. A practical implementation of such a detector
could suffer from: i) unwanted non-GW active resonant modes in the
frequency region of interest, ii) additional thermal noise, due to
the low frequency contribution of the high frequency non GW
sensitive modes \cite{conti2}. These difficulties may be overcome
by introducing a \textit{geometrically selective} readout, capable
of specifically selecting the contribution to the signal from all
the GW sensitive modes. We have found also an additional bonus:
the scheme is capable to give an effective reduction of the
back-action noise within the bandwidth. The overall effect is a
significant enhancement in the sensitivity.

In a \textquotedblleft dual{\textquotedblright} detector  one
would measure the differential displacement, driven by the GW, of
the nearly faced surfaces of two concentric massive bodies,
mechanically resonating at different frequencies \cite{cerdonio1}.
In this scheme the centers of mass of the two bodies coincide and
then remain mutually at rest while the masses resonate. The
differential displacement is then only due to the internal
vibrational modes, while the centers of mass provide the rest
frame for the measurement. This design allows the use of wide
bandwidth (non resonant) readouts, evolution in concept and in
technology of the resonant readouts used for bar detectors
\cite{capacitivo}.

\begin{figure}[t!]
\includegraphics[width=8.6cm,height=6.5cm]{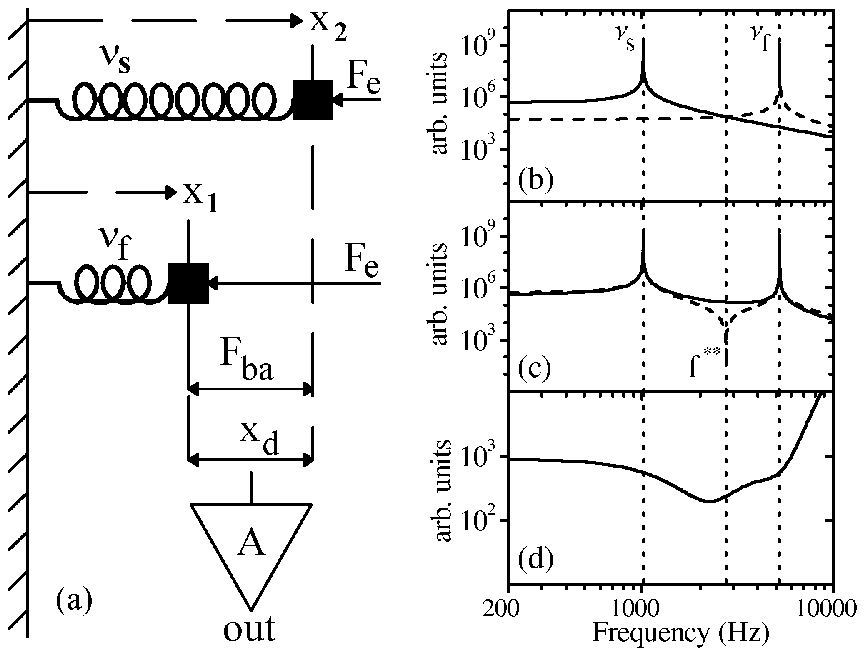}
\caption{\label{fig:paper0} (a) One-dimensional \textquotedblleft
dual{\textquotedblright} detector: the same force $F_e$ is
measured by the relative displacement $x_d$ of two resonators. (b)
Transfer functions of the slow resonator (continuous line) and of
the fast resonator (dashed line). (c) Dual detector transfer
functions: signal $H_{F_e}={x_d}/{F_e}$ (continuous line),
back-action $H_{ba}={x_d}/{F_{ba}}$ (dashed line). (d) Wideband
optimized noise. }
\end{figure}

A basic dual detector can be represented as a simple one
dimensional system [Fig.~\ref{fig:paper0}(a)], where a force
$F_e$, acting on two different mechanical resonators, is evaluated
by a differential measurement of their positions $x_1$ and $x_2$.
In the frequency region between the two resonant frequencies,
$F_e$ drives the slow resonator above its resonance $\nu _s$ and
the fast one below its resonance $\nu _f$. The responses of the
two resonators are then out of phase by $\pi$ radians and
therefore the differential motion $x_d$ results in a signal
enhancement over the single oscillator responses
[Fig.~\ref{fig:paper0}(b)], as shown by the transfer function
$H_{F_e}={x_d}/{F_e}$ [Fig.~\ref{fig:paper0}(c)].

Such a scheme in addition leads to a reduction of the back-action
noise, just within the detector bandwidth of interest. In fact the
system response to the back-action force, that it is coherently
applied with \textit{opposite} direction on the two masses, is
nearly exactly in phase and the consequent differential
displacement is highly depressed at a frequency $f^{**}$, as shown
in the transfer function $H_{ba}={x_d}/{F_{ba}}$ (Fig.
\ref{fig:paper0}c). We measure $x_d$ with a displacement
amplifier, described by its additive displacement and back-action
force noises, with white power spectra $S_{XX}(\omega)=S_{xx}$ and
$S_{FF}(\omega)=S_{ff}$, so that the total displacement noise is
$S_{xx}+|H_{ba}(\omega)|^2S_{ff}$. The noise power spectrum on the
measurement of $F_e$, due to the amplifier system, is then:
\begin{equation}
\label{eq:genhhpowerdef}
 S_{F_e}(\omega)=\big({S_{xx}+|H_{ba}(\omega)|^2 S_{ff}}\big)/{|H_{F_e}(\omega)|^2}\,.
\end{equation}
If we take as reference an operation at the Standard Quantum Limit
(SQL), we may consider $S_{xx}S_{ff}=\frac{\hbar^2}{4}$, and the
noise figure can be optimized by adjusting the ratio
$S_{xx}/S_{ff}$. In a wide bandwidth detection strategy $S_{xx}$
and $S_{ff}$ must be balanced to give the lowest noise within the
bandwidth. In doing so we profit by the subtraction effect in the
back-action noise transfer function $H_{ba}$, and obtain a dip at
the frequency $f^{**}$, as shown in Fig.~\ref{fig:paper0}(d). We
finally notice that, to fully exploit the back-action reduction
features, $f^{**}$ should be placed, by a proper choice of the
system parameters, amid the oscillator frequencies.

In the case of a  three-dimensional body,  the dynamics of elastic
deformations is given as the superposition of the dynamics of an
almost infinite number of normal modes of vibration \cite{Love}.
An obvious way to preserve the convenient features for signal and
back action noise outlined above, is to bring the real system to
be as close as possible to the idealized two modes system. In fact
when only  the first quadrupolar mode can be considered for each
body, the response to a GW of such a system can again be described
by the simple one-dimensional model. This can be accomplished with
a novel selective readout we propose here, capable of rejecting a
large number of normal modes on the basis of their symmetry. This
\textit{geometrically based mode selection} senses the surface
position of a body on specific regions, so that the related
deformations can be combined with a weight/sign properly chosen to
optimize the total response to normal modes of quadrupolar
symmetry. Such a strategy to select specific vibrational modes and
to reject a class of unwanted modes is conceptually different from
the strategy now employed in GW acoustic detectors. Resonant bar
detectors \cite{bars} and spherical detectors
\cite{hamilton,sfera} reconstruct the amplitude of the normal
modes excited by the GWs by the use of resonant displacement
readouts coupled to the modes. With a proper choice of the
read-out surfaces, the resonant scheme is not sensitive to the
thermal noise of out of resonance modes and gives an efficient
\textit{frequency based mode selection}. However this feature
necessarily limits the detector bandwidth, due to the thermal
noise contribution of the resonant transducer, as in every
resonant readout scheme. By contrast the \textit{geometrically
based mode selection} selects gravitationally sensitive normal
modes by means of their geometrical characteristics, and shows a
significant rejection of non quadrupolar modes without affecting
the detector bandwidth. For this reason it can be effectively
applied to a \textquotedblleft dual{\textquotedblright} wideband
detector.

\begin{figure}[t!]
\includegraphics[width=8.6cm,height=4.1cm]{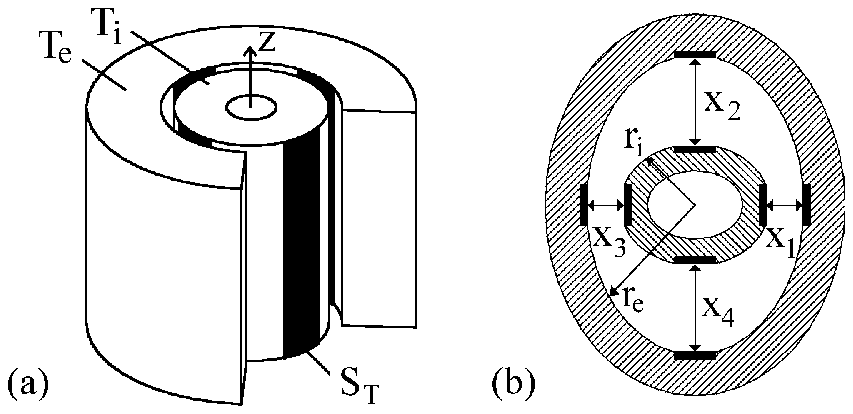}
\caption{\label{fig:paper1}  (a) The two concentric torii,
$T_e,\;T_i$  are made of  materials of density $\rho_e,\;\rho_i$
and have the same height. The inner torus may also have null
internal radius and be reduced to a cylinder. The relative
distance between the two bodies is measured in 4 regions (in
black),  each of area $S_T$, for the whole torus height. (b)
Section of the detector showing the signal enhancement obtained
when a GW signal drives the external torus above resonance and the
internal torus below resonance. The difference
$X_d=x_1+x_3-x_2-x_4$, proportional to the GW strength, is not
dependent on a number of non GW sensitive modes.}
\end{figure}

Without loss of generality, we apply these new concepts to a dual
torus detector [Fig.~\ref{fig:paper1}(a)],  a convenient geometry
where the advantages of the proposed scheme can be fully
exploited. In the dual torus we average the differential
displacement in 4 distinct areas $x_{1..4}$
[Fig.~\ref{fig:paper1}(b)] and  combine them to obtain
$X_d=x_1+x_3-x_2-x_4$. The detector displays its maximal
sensitivity when the GW propagates along the $z$ axis, the
symmetry axis of the system. The corresponding force does not
depend on \textit{z} and the system response can be well described
by plane strain solutions, where the displacements are functions
of \textit{x} and \textit{y} only and the displacement along
\textit{z} vanishes. All the components of the internal stress are
also independent by \textit{z}. In this case the displacement
normal modes of a single torus, eigenfunctions of the free body
dynamic equations, are functions of the kind:
\begin{eqnarray}
\mathbf{w}^+_{a,\,n}(\mathbf{r})&=&f_{a,\,n}\,\cos\,(a\,\theta)\:\mathbf{i}_r\:+\:g\,_{a,\,n}
\,\sin \,( a \,\theta )\:\mathbf{i}_\theta \label{twinmodes+} \nonumber\\
\mathbf{w}^\times_{a,\,n}(\mathbf{r})&=&-f_{a,\,n}\,\sin\,(a\,\theta)\,\mathbf{i}_r\:
+\:g\,_{a,\,n}  \cos  \,(a \,\theta)\,\mathbf{i}_\theta\,,
\label{twinmodesx}\nonumber
\end{eqnarray}
where the functions $f_{a,\,n}, g\,_{a,\,n}$ are linear
combinations of Bessel functions of the coordinate $r$, with
coefficients given by the boundary conditions. The integer $a$
represents the angular symmetry of the mode, while n identifies
the mode order within the angular family. The orthogonal
displacement fields
$\mathbf{w}^+_{a,\,n},\mathbf{w}^{\times}_{a,\,n}$  represent the
same radial distribution of the deformation, mutually rotated by
$\frac{\pi}{2a}$: for this reason they share the same eigenvalue
$\omega_{a,\,n}$, called resonant frequency of the mode. Any
displacement $\mathbf{u}$ may be written as linear superposition
of these basis functions:
\begin{equation}
 \mathbf{u}(\mathbf{r},t)= \sum_{s,\,a,\,n}
 \mathbf{w}^s_{a,\,n}(\mathbf{r})\,q^s_{a,\,n}(t)\,
 ,\label{sumsolution}
\end{equation}
where the time dependent coefficients are determined by the force
acting on the body and $s=+,\times$.

\begin{figure}[t]
\includegraphics[width=8.6cm,height=6.1cm]{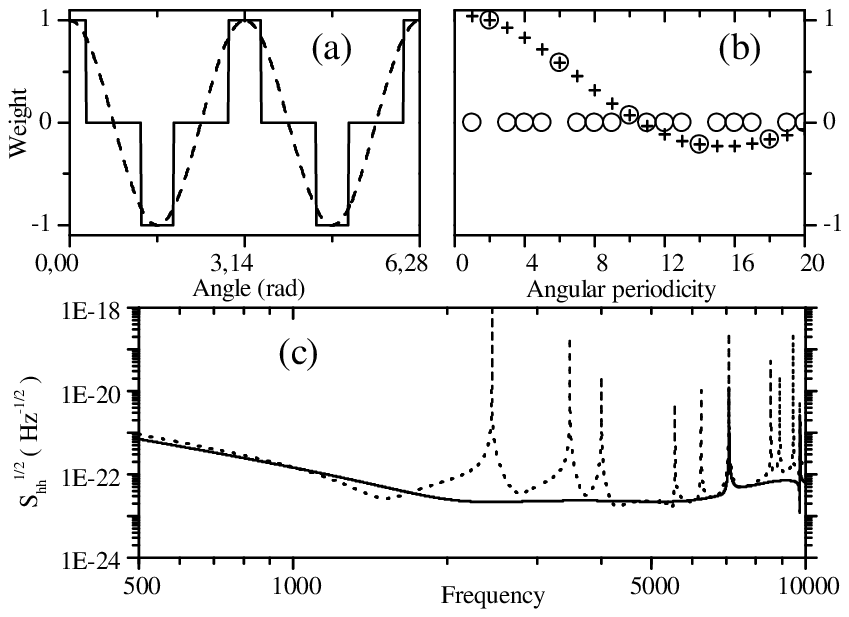}
\caption{\label{fig:paper2} (a) Angular dependence of the weight
function $\mathbf{P}_4$ (continuous line) and of the displacement
induced by a $\mathbf{w}^+$ mode (dashed line) with symmetry
$a=2$; the mode contribution is proportional to the integral of
the product of these two functions. (b) Normalized angular
contribution of normal modes as a function of $a$ evaluated for
the weight functions $\mathbf{P}_4$ (hollow symbols) and
$\mathbf{P}_1$ (crosses). (c) Predicted sensitivity of a Mo dual
torus detector (same as Fig. \ref{fig:paper5}) with the selective
read-out $\mathbf{P}_4$ (continuous line) and with the standard
read-out $\mathbf{P}_1$ (dotted line).}
\end{figure}

A $+$ polarized GW propagating along the $z$-axis applies on the
mass of density $\rho$ the force density
$F_{gw}(t)\,\mathbf{G}_{gw}(\mathbf{r})$:
\begin{eqnarray}
F_{gw}(t)&=&\frac{1}{2} \,\rho \,\ddot{h}(t) \nonumber\\
\mathbf{G}_{gw}(\mathbf{r})&=& r \,\cos\,(\,2\,\theta \,)
\,\mathbf{i}_r - r\,\sin\,(\,2\,\theta \,)\, \mathbf{i}_\theta\,.
\label{gwaveplus}
\end{eqnarray}
For symmetry reasons this force can only excite $\mathbf{w}^+$
quadrupolar  ($a=2$) modes. A \textquotedblleft weight
function{\textquotedblright} approach to the
problem will give the mathematical framework to %
study the selective read-out, which implements the difference
$X_d=x_1+x_3-x_2-x_4$ [Fig.~\ref{fig:paper1}(b)]. If
$\mathbf{u}_e$ and $\mathbf{u}_i$ are the displacements of the
torus $T_e$ and $T_i$ \cite{independent}, we define the observable
physical quantity of the system as:
\begin{equation}
\label{observablediff} X_d(t)= \,\int\,[ \,
\mathbf{u}_e(\mathbf{r},t)+ \mathbf{u}_i(\mathbf{r},t)\,] \,\cdot
\,\frac{\mathbf{P}_4(\mathbf{r})}{P_{N}}\, \textit{d}\, V\;,
\end{equation}
where the  \textquotedblleft selective{\textquotedblright}
measurement strategy and detection scheme is implemented by the
weight function
$\mathbf{P}_4(\mathbf{r})=P^r_{4}(r)\,P^{\theta}_{4}(\theta)\,\mathbf{i}_r$,
where:
\begin{eqnarray}
P^r_{4}&=& \delta(r-r_{e})-\delta(r-r_{i})\nonumber\\
P^{\theta}_{4}&=&\sum_{m=0}^1\, \sum_{n=0}^4 (-1)^{n+m}
\,\Theta\,[\,\theta+(-1)^{m}\alpha -n\frac{\pi}{2}\, ] \,\:
\label{eqn:Pweight}
\end{eqnarray}
and $\Theta (x)$ represents the unit step function. The
normalization is $P_{N}=S_T$,  area of one sampling region
[Fig.~\ref{fig:paper1}(a)]. Here the term $P^r_{4}$ gives the
requested displacement difference $r_e-r_i$
[Fig.~\ref{fig:paper1}(b)], while $P^{\theta}_{4}$ reduces the
angular integral over 4 distinct regions, $2\alpha$ wide, centered
around $\theta=0,\frac{\pi}{2},\pi,\frac{3\pi}{2}$
[Fig.~\ref{fig:paper2}(a)]. A value $\alpha=0.3$ rad is assumed to
perform the following calculations. For comparison we consider a
non selective transducer system, which senses the displacement
over a single area, centered for example at $\theta=0$. Its weight
function $\mathbf{P}_1$ has angular component
$P^{\theta}_{1}=\Theta\,(\,\theta+\alpha)-\Theta\,(\,\theta-\alpha
)+\Theta\,(\,\theta+\alpha-2\pi)-\Theta\,(\,\theta-\alpha-2\pi )
$, while the radial dependence and the normalization remain the
same as $\mathbf{P}_4$.

When we evaluate the observable Eq. (\ref{observablediff}), by
using the expansion Eq. (\ref{sumsolution}) and  the weight
functions $\mathbf{P}_4$ or $\mathbf{P}_1$, each
$\mathbf{w}^+_{a,\,n}$ mode contribution depends on its angular
symmetry $a$. As shown in Fig.~\ref{fig:paper2}(b), in both cases
the modes contribution oscillates and rapidly decreases, due to
the averaging over the area $S_T$. But in the case of
$\mathbf{P}_4$ only the symmetry values $a=2 + 4\,k$, with $k$ non
negative integer, give non null contributions to our observable
Eq. (\ref{observablediff}). To summarize, the quadrupolar modes
family ($k=0$) contribution is essentially preserved, many mode
families are rejected ($a\neq 2+4k$, $k\geq 0$) and the residual
higher order families ($a=2 + 4\,k$, $k>0$) give a reduced
contribution.

We point out that the normal modes $\mathbf{w}^\times$,
proportional to $\sin\,(a \,\theta)$, and excited by
$\times$-polarized gravitational waves, are rejected by
$\mathbf{P}_4(\mathbf{r})$, for every value of $a$. When a second
transducer system, identical but rotated by $\pi/4$, is employed
to detect these $\mathbf{w}^\times$ modes, any $z$-axis
propagating GW can be fully characterized in terms of intensity
and polarization.

The detector sensitivity  can be evaluated by the transfer
function $T_{X_d} \equiv \widetilde{X}_{d}(\omega) /
\widetilde{F}(\omega)$, which gives the observable $X_d$ induced
 by a generic driving force
density $F(t)\,\mathbf{G}(\mathbf{r})$ \cite{forcecondition}. We
call ($\mathbf{w}^s_{a,\,n},\, \omega_{\,a,\,n}$) and
($\mathbf{v}^s_{a,\,n},\, \varpi_{\,a,\,n}$)  the normal modes and
eigenfrequencies of the torus $T_e$ and $T_i$, while the loss
angles ($\phi , \, \psi )$, inversely proportional to the mode
quality factor $Q$, describe the dissipation. We have:
\begin{widetext}
\begin{eqnarray} \label{difftransfergeneric}
T_{X_d}(\omega) =\sum_{s,\,a,\,n}
\,\frac{\int\mathbf{G}(\mathbf{r}) \cdot
\mathbf{w}^s_{a,\,n}(\mathbf{r}) \textit{d}V\; \int
\mathbf{w}^s_{a,\,n}(\mathbf{r}) \cdot
\frac{\mathbf{P}(\mathbf{r})}{P_{N}} \textit{d}
V}{\rho_e\,[(\omega^2_{\,a,\,n}-\,\omega^2\,)+ i
\,\omega^2_{\,a,\,n} \phi_{\,a,\,n}(\omega)]}+
     \frac{\int\mathbf{G}(\mathbf{r})\cdot\mathbf{v}^s_{a,\,n}(\mathbf{r})
\textit{d}V\; \int \mathbf{v}^s_{a,\,n}(\mathbf{r})\cdot
\frac{\mathbf{P}(\mathbf{r})}{P_{N}} \textit{d}
V}{\rho_i\,[(\varpi^2_{\,a,\,n}-\,\omega^2\,)+ i
\,\varpi^2_{\,a,\,n} \psi_{\,a,\,n}(\omega)]}\,\:,
\end{eqnarray}
\end{widetext}
where $ \mathbf{P}(\mathbf{r}) $ can be $ \mathbf{P}_4(\mathbf{r})
$ or $ \mathbf{P}_1(\mathbf{r})$.
 If $F(t)$ and
$\mathbf{G}(\mathbf{r})$ are given by Eq. (\ref{gwaveplus}), by
Eq. (\ref{difftransfergeneric}) we calculate the system transfer
function to a GW, defined as $H_{gw}(\omega)\equiv
{\widetilde{X}_{d}(\omega)} /{\widetilde{h}(\omega)}$. In the
simple case $\rho_e=\rho_i=\rho$, we have:
\begin{equation}
\label{eq:Hgw} H_{gw}(\omega) = -( {\rho \, \omega^2}/{2})\;
T_{X_d}(\omega)\, \big|
_{\mathbf{G}(\mathbf{r})=\mathbf{G}_{gw}(\mathbf{r})}\;,
\end{equation}
while the more general case is straightforward. The read-out
back-action force is applied in the sensing areas with the same
intensity but opposite sign for the two bodies, so that its
spatial density is given by
$\frac{\mathbf{P}(\mathbf{r})}{P_{N}}$. The corresponding system
transfer function is then:
\begin{equation}
\label{eq:Hba}
 H_{ba}(\omega)\equiv T_{X_d}(\omega)\big|
_{\mathbf{G}(\mathbf{r})=\mathbf{P}(\mathbf{r})/P_N}\;.
\end{equation}

\begin{figure}[h!]
\includegraphics[width=8.6cm,height=4.2cm]{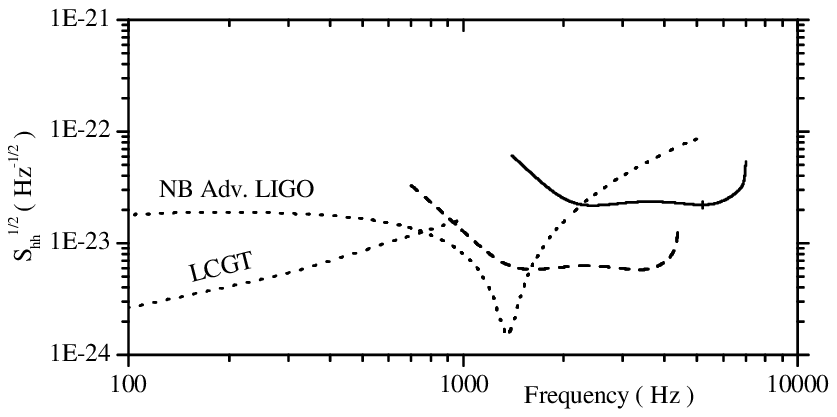}
\caption{\label{fig:paper5}Predicted spectral strain SQL
sensitivities of two different dual detector configurations. The
predicted SQL sensitivities of two advanced interferometric
detectors are also shown (dotted lines): LCGT and a narrow band
design of Advanced LIGO. Continuous line: Mo dual detector, inner
cylinder radius $0.25$ m, outer torus internal-external radius
$0.26\,$-$\,0.47$ m, height $2.35$ m, weight  4.8 + 11.6 tons,
fundamental quadrupolar modes 1012 Hz and 5189 Hz, amplifier noise
$S_{xx}=5 \times 10^{-46}$ m$^2$/Hz, $Q/T>2\times 10^8$ K$^{-1}$.
Dashed line: SiC detector, inner cylinder radius $0.82$ m, outer
torus internal-external radius $0.83\,$-$\,1.44$ m, height $3$ m,
weight 20.5 - 41.7 tons, $S_{xx}=3 \times 10^{-46}$ m$^2$/Hz,
${Q}/{T}>2\times 10^8$ K$^{-1}$.}
\end{figure}

In close analogy with Eq. (\ref{eq:genhhpowerdef}), the transfer
functions Eqs. \ref{eq:Hgw} and \ref{eq:Hba} give the detector
sensitivity to GW as:
$S_{hh}(\omega)=\big({S_{xx}+|H_{ba}(\omega)|^2
S_{ff}}\big)/{|H_{gw}(\omega)|^2}$. In Fig.~\ref{fig:paper2}(c) we
compare the sensitivities of a detector, evaluated  for the
selective readout $\mathbf{P}_4$ and for the single area read-out
$\mathbf{P}_1$.

The main effect of the selective readout is to cancel out both
thermal and back-action noise contributions due to the non
quadrupolar modes, so that a flat response is obtained in the full
bandwidth of few kHz and the back action reduction feature is
exploited. The detector dimensions are properly chosen to profit
by the GW sensitivity of the second order quadrupolar mode of the
outer torus. We also notice that the use of large read-out sensing
area highly reduces the thermal noise due to the cumulative effect
of all the normal modes \cite{gillespie}. In fact, as the shorter
wavelength modes are averaged out, a very good transfer function
convergence may be obtained by adding less than 100 modes.

Finite element analysis demonstrates the selective read-out
rejection capabilities also for many classes of 3-dimensional
vibrational modes. The sensitivity enhancement over the standard
readout scheme is then not limited to our plane strain
approximation, and the dual torus configuration could be evolved
in a complete detector. A practical readout configuration is a
series of 4 capacitive transducers, gradiometrically connected and
sensed by a single SQUID amplifier. This implements the selective
scheme $\mathbf{P}_4$ and the consequent back-action reduction.
The needed sensitivity could be reached as recent progress show
that the SQUID amplifiers are approaching the quantum limit, now
also in the necessary strong coupling configuration
\cite{vinante}, and that the electric polarization field can be
increased up to the material intrinsic limitations
\cite{kobayashi}.

To evaluate the limits of our design, we can evaluate the
sensitivity at the SQL of some practical configurations of
detector material and geometry. As usual a low dissipation
material is required to reduce the effect of the thermal noise.
Molybdenum represents an interesting choice, as it shows high
cross-section for GWs and its mechanical dissipation was
investigated at low temperature giving $Q/T>2\times 10^8$ K$^{-1}$
for acoustic normal modes \cite{duffyMo}. In
Figs.~\ref{fig:paper2} and \ref{fig:paper5} is shown the SQL
sensitivity of a Mo detector with dimensions within the present
technological production capabilities. In Fig.~\ref{fig:paper5} we
also show the SQL of a detector made of SiC, a ceramic material
currently used to produce large mirrors or structures \cite{SiC},
with mechanical and thermal properties of interest here but not
jet characterized in terms of low temperature mechanical
dissipation.

The selective readout scheme applied to the dual concept allows
the design of detectors tailored for relatively high frequency GW
and with very few spurious modes within their wide bandwidth.
These features could make the dual torus detector complementary to
advanced interferometric detectors, as shown in
Fig.~\ref{fig:paper5} by the comparison with the expected
sensitivities of LCGT \cite{kuroda} and one of the possible
setting of Advanced LIGO in narrow band operation \cite{ligo2}.


\begin{thebibliography}{<20>}
\bibitem{cerdonio1} M. Cerdonio, \textit{et al}., Phys. Rev. Lett. \textbf{87} 031101
(2001).
\bibitem{sorgenti} K. Thorne in \textit{300 Years of Gravitation},
S.W. Hawking and W. Israel eds. (Cambridge University Press, N.Y.
1987)
\bibitem{conti2} L. Conti \textit{et al}. Class. Quant. Grav. \textbf{19} 2013
(2002).
\bibitem{capacitivo} A. Marin \textit{et al}., Class. Quant. Grav. \textbf{19} 1991
(2002).
\bibitem{Love} A.E.H. Love, \textit{A treatise on the mathematical theory
of elasticity} (Dover, New York, 1944).
\bibitem{bars} Z.A. Allen \textit{et al.}, Phys. Rev. Lett. {\bf 85}, 5046 (2000)
\bibitem{hamilton}S.M. Merkowitz \textit{et al.}, Phys. Rev. D \textbf{56}, 7513
(1997).
\bibitem{sfera} J.A. Lobo, Mon. Not. R. Astron. Soc. \textbf{316} 173
(2000).
\bibitem{independent} We assume that the solutions $\textbf{u}_e$ and $\textbf{u}_i$ are
independent.
\bibitem{forcecondition} In our approximation
this force must be null, with its derivatives, along the $z-$axis.
This condition is fulfilled by the considered gw force and
back-action force.
\bibitem{gillespie} A. Gillespie and F. Raab, Phys. Rev. D \textbf{52}, 577 (1995).
\bibitem{vinante} A. Vinante \textit{et al}. Appl. Phys. Lett. {\bf 79}, 2597
(2001).
\bibitem{kobayashi} S. Kobayashi, IEEE Trans. on Diel. Elec.
Ins. \textbf{4}, 841 (1997).
\bibitem{duffyMo} W. Duffy, Jr., J.
Appl. Phys. \textbf{72}, 5628 (1992).
\bibitem{SiC} B. Harnisch \textit{et al}., ESA bulletin \textbf{95}
(1998).
\bibitem{kuroda}K. Kuroda \textit{et al.}, Class. Quantum Grav. \textbf{19}, 1237 (2002)
\bibitem{ligo2}P. Fritschel in \textit{Gravitational-Wave Detection},
P. Saulson and M. Cruise eds., Proc. SPIE \textbf{4856} (2003).

\end{thebibliography}
\end{document}